\documentclass[letterpaper,twocolumn,10pt]{article}
\PassOptionsToPackage{hyphens}{url}
\usepackage{usenix-2020-09}

\def\isanonymous{0}
\def\isfullversion{1}

\usepackage[l2tabu,orthodox]{nag}

\usepackage{ifxetex}

\usepackage{microtype}

\usepackage{hyphenat}
\usepackage{ifthen}
\newcommand{\anonymous}[2]{%
\ifthenelse{\equal{\isanonymous}{1}}%
{#1}%
{#2}%
}

\newcommand{\fullversion}[2]{%
\ifthenelse{\equal{\isfullversion}{1}}%
{#1}%
{#2}%
}

\usepackage{xcolor}
\definecolor{oxygenorange}{HTML}{FFDD00}
\usepackage[color=oxygenorange]{todonotes}

\ifthenelse{\equal{\isanonymous}{1}}
{
  \newcommand{\malb}[2][]{}
  \newcommand{\rikke}[2][]{}
  \newcommand{\jorge}[2][]{}
  \newcommand{\lenka}[2][]{}
  \renewcommand{\todo}[2][]{}
}{
  \newcommand{\malb}[2][inline]{\todo[#1]{\textbf{malb:} #2}\xspace}
  \newcommand{\rikke}[2][inline]{\todo[#1]{\textbf{rikke:} #2}\xspace}
  \newcommand{\jorge}[2][inline]{\todo[#1]{\textbf{jorge:} #2}\xspace}
  \newcommand{\lenka}[2][inline]{\todo[#1]{\textbf{lenka:} #2}\xspace}
}

\usepackage{amsmath,amsfonts}  
\usepackage{xspace}

\usepackage[lambda,landau,operators,probability,sets,logic,complexity,asymptotics]{cryptocode}

\usepackage{booktabs}  
\usepackage{comment}
\usepackage{enumitem}
\usepackage{longtable}
\usepackage{navigator}

\usepackage{subcaption}   
\usepackage{tikz,pgfplots,pgfplotstable}
\usepgfplotslibrary{statistics}
\usetikzlibrary{calc}
\usetikzlibrary{arrows}
\usetikzlibrary{positioning}

\pgfplotsset{
	tick label style={font=\small},
	label style={font=\small},
	legend style={font=\small, cells={anchor=west}}
}

\definecolor{DarkPurple}{HTML}{332288}
\definecolor{DarkBlue}{HTML}{6699CC}
\definecolor{LightBlue}{HTML}{88CCEE}
\definecolor{DarkGreen}{HTML}{117733}
\definecolor{DarkRed}{HTML}{661100}
\definecolor{LightRed}{HTML}{CC6677}
\definecolor{LightPink}{HTML}{AA4466}
\definecolor{DarkPink}{HTML}{882255}
\definecolor{LightPurple}{HTML}{AA4499}

\definecolor{DarkBrown}{HTML}{604c38}
\definecolor{DarkTeal}{HTML}{23373b}
\definecolor{LightBrown}{HTML}{EB811B}
\definecolor{LightGreen}{HTML}{14B03D}

\usepackage{listings}
\lstdefinelanguage{Sage}[]{Python}{morekeywords={True,False,sage,cdef,cpdef,ctypedef,self},sensitive=true}
\begin{document}

\title{Collective Information Security in Large-Scale Urban Protests: \\ the Case of Hong Kong}
\anonymous{
  \author{}
}{
  \author{
    {\rm Martin R. Albrecht}\\
    Royal Holloway, University of London\\
    martin.albrecht@rhul.ac.uk
	\and
    {\rm Jorge Blasco}\\
    Royal Holloway, University of London\\
    jorge.blascoalis@rhul.ac.uk
	\and
    {\rm Rikke Bjerg Jensen}\\
    Royal Holloway, University of London\\
    rikke.jensen@rhul.ac.uk
	\and
    {\rm Lenka Mareková}\\
    Royal Holloway, University of London\\
    lenka.marekova.2018@rhul.ac.uk
  }
      
}

\fullversion{}{
  \pagestyle{empty}
  \thispagestyle{empty}
}

\maketitle
\begin{abstract}
  The Anti-Extradition Law Amendment Bill protests in Hong Kong present a rich context for exploring information security practices among protesters due to their large-scale urban setting and highly digitalised nature. We conducted in-depth, semi-structured interviews with 11 participants of these protests. Research findings reveal how protesters favoured Telegram and relied on its security for internal communication and organisation of on-the-ground collective action; were organised in small private groups and large public groups to enable collective action; adopted tactics and technologies that enable pseudonymity; and developed a variety of strategies to detect compromises and to achieve forms of forward secrecy and post-compromise security when group members were (presumed) arrested. We further show how group administrators had assumed the roles of leaders in these `leaderless' protests and were critical to collective protest efforts.
\end{abstract}

\section{Introduction}\label{sec:introduction}

Large-scale urban protests offer a rich environment to study information security needs and practices among groups of higher-risk users by relying on a diverse set of digital communication platforms, strategies and tactics, and by their sheer size. In this work, we study the Anti-Extradition Law Amendment Bill (Anti-ELAB) protests in Hong Kong, where most activities and interactions map onto some form of digital communication. The use of different communication platforms as an integral part of the protests has already been documented in various media reports, including: large chat groups on platforms such as Telegram, protest-specific forums on the Reddit-like platform LIHKG, practices of doxxing as well as live protest maps such as HKmap.live to identify police positions~\cite{press:HK:abacus,press:HK:HKFP,press:HK:NYT, press:HK:times}. Recent scholarship has also highlighted the significance of digital technology to the Anti-ELAB protests. For example, ``novel uses'' of communication technology by Anti-ELAB protesters led them to form ad hoc and networked ``pop-up'' protests, creating a new form of a ``smart mob'' facilitated by digital technology~\cite{SMS:Ting20}. Platforms such as Telegram and LIHKG worked to mobilise and establish a sense of community among young activists~\cite{SP:SKu20} and created a ``symbiotic network'' of protesters~\cite{CHI:KowNarChe20}. Social media was used to maintain ``protest potential'' over time~\cite{IJC:LeeChaChe20}.

To design and build secure communication technologies that meet the needs of participants in large-scale protest movements, it is critical that designers and technologists understand protesters' specific security concerns, notions, practices and perceptions. There is also a need to understand the existing use of secure and appropriation of insecure communication tools within such protest groups, where they fail and where they succeed. Existing qualitative studies have explored security practices of different groups of higher-risk users, e.g.~\cite{ErmHalMus17,HalErmMus18,SP:DSKB21,USENIX:MCHR15,PETS:McgRoeCai16,CHI:GMSMTS18,CHI:ColJen19,CHI:ColJenTal18,SP:SLIRK18,CHI:LHKZSH20}, but none to our knowledge have studied such practices within large-scale urban protests.

The Anti-ELAB protests, while specific in nature like any other local protest movement, provide ample material for a case study. This is not only for the features already outlined above -- urban, large-scale, digitalised -- but also because of the place these protests take in the imagination of protest movements across the globe. The perceived analogue and digital tactics developed in Hong Kong have been imitated by protesters elsewhere, often with a direct reference, see e.g.~\cite{press:IN:scmp,Chuang20,press:qz}. 

\subsubsection*{Contributions}\label{sec:contributions}

We develop a grounded understanding of (perceived and actual) security needs and practices among Anti-ELAB protesters through in-depth, semi-structured interviews with 11 participants from Hong Kong. Through an inductive analysis of these interviews, research findings were synthesised into five main categories. We outline these in Section~\ref{sec:findings} -- the tools used by Anti-ELAB protesters and the reasons for their adoption (Section~\ref{sec:tools}), the role these tools play for the organisation of these protests (Section~\ref{sec:social-organisation}), the tactics used to detect and mitigate compromises through arrests (Section~\ref{sec:forward-secrecy}), the practices adopted to work around limitations of the tools relied upon (Section~\ref{sec:limitations}) and the routes and negotiations through which protesters arrive at their understanding and practice of security (Section~\ref{sec:ideas-security}) -- before bringing these into conversation with information security scholarship in Section~\ref{sec:discussion}, where we also identify open research questions, and concluding in Section~\ref{sec:conclusion}.

\section{Related Work}\label{sec:related-work}

We position our research within studies on digital communication technology use by participants of large protest movements, including existing work on the Anti-ELAB protests to establish pre-existing understanding of their technology use, as well as scholarly work on higher-risk users.

\subsection{Large-scale protests and digital communication}\label{sec:large-scale-protests}

The importance of digital communication technology in large-scale protests is well documented in the social science literature, focusing in particular on the significant contribution of social media platforms to the mobilisation of social movements~\cite{Castells12,Coopman11,DenLei15,Ems14,LeeCha10,MeNePo18,MJHY15,Shirky2011,VanVan10}. They also highlight the critical role that digital media play in the organisation and coordination of large-scale protests, e.g.~Occupy Wall Street and the Arab Spring~\cite{AlsGuv15,Fuchs14,HowHus13,Kavada15,Nielsen13,Tremayne14,TufWil12}. Yet, there is consensus in the literature that while the ability to form online networks can support mobilisation and organisation efforts, it is neither the sole driver nor the underlying cause. 

Scholars also note how digital communication technology enables new networks and movement formations. For example, Bennett and Segerberg~\cite{BenSeg12} describe a form of protest movements not reliant on resourceful organisations, but driven by personal online content and communications -- what they call ``connective action''. Others, e.g.~\cite{Castells12,Kavada15,MJHY15}, highlight how digital technology enables the formation of decentralised networks among groups in different locations, through collective action. These movements are able to attract large numbers of participants, partly because they are supported by digital infrastructures~\cite{LeeCha16}. Studies have also suggested that people ``self-mobilise'' online before taking part in protests~\cite{Harlow12,Lee15,SPLT11}.
Finally, digital technologies are often used to facilitate on-the-ground  organisation, information sharing and communication between protesters -- what Trer{\'e}~\cite{Trere15,Trere20} calls ``backstage activism''. 

\subsubsection*{Messaging applications}
Some studies explore the use of messaging applications in distinct resistance movements and protest environments. For example, Uwalaka et al.~\cite{UwRiWa18} considered the use of WhatsApp in the 2012 Occupy Nigeria protest, Gil de Z{\'u}{\~n}iga et al.~\cite{GiArAl19} and Valeriani and Vaccari~\cite{ValVac18} studied messaging applications in activism and political organisations, while Trer{\'e}~\cite{Trere20} showed how WhatsApp is used for everyday activities and organisation by protesters in Spain and Mexico. Similarly, Haciyakupoglu and Zhang~\cite{HacZha15} found that in the Gezi Protests in Turkey protesters relied especially on WhatsApp to circulate information within the protest area.
Messaging applications have also been linked to the spreading of rumours and incitement to violence. For example,
Mukherjee~\cite{Mukherjee20} explored the use of WhatsApp in mob lynchings in India and Arun~\cite{Arun19} linked the spreading of rumours via Whats\-App to them.
Tracking and hacking on digital communication platforms are also used by private and state actors to counter opposition movements and to suppress dissent~\cite{intercept:gulf,citizenlab:nso}.

While such prior works do not consider (information) security in particular, they provide broader context and in some cases surface security-related findings. For example, the importance of trust in information, technology and social media networks is explored in~\cite{HacZha15,LeeCha16} and Tsui~\cite{Tsui15} studies digital technology use and protection from state surveillance efforts, while Sowers and Toensing~\cite{SowToe12} engage with wider security concerns such as threats to protesters from authoritarian and violent regimes.

\subsection{Anti-ELAB protests}
The protests responded to the Hong Kong Government's attempt to pass an Extradition Law Amendment Bill~\cite{CAS:Lee20,CR:LYTC19}. Hundreds of thousands of people took to the streets, where networked groups of protesters organised mass rallies and strikes, boycotted pro-Beijing businesses, barricaded streets, stormed public buildings including the Legislative Council Complex, occupied traffic hubs and seized university campuses~\cite{IJS:Holbig20}. Recent studies have emphasised the centrality of digital and mobile communication technology to facilitate these large, dynamic and highly mobile protest activities; with tactics often referred to as ``be water'' and ``blossom everywhere''~\cite{press:HK:Hale}. Such tactics meant that the protests emerged from the ground up among activist networks in a nonhierarchical, diversified fashion, relying on spontaneous initiatives rather than top-down leadership and organisation. In general, this served two purposes. While it provided protection from prosecution of individual protesters and police detection, it gave rise to fluid, horizontal communication within and between dispersed groups of protesters~\cite{IJS:Holbig20}. These tactics were partly rooted in protesters' experiences from the 2014 Umbrella Movement in Hong Kong, where high-profiled protesters were arrested and imprisoned, and which were also supported by digital modes of participation that enabled, for example, real-time coordination of ``improvisatory acts''~\cite{LeeCha16}.

The Anti-ELAB protests are widely considered to have been ``innovative'' in their tactics, particularly the interaction between ``front line'' protesters and others. A ``frontliner'', roughly, is someone engaging in activities that risk direct confrontation with law enforcement~\cite{Chuang20}. An example of a collaboration between ``frontliners'' and others are ride sharing schemes where car owners picked up ``frontliners'' to transport them out of the protest area because public transport was deemed unsafe or shut down~\cite{press:hk-ride-sharing-reuters}. These schemes were run via public online groups that connected protesters with drivers.

Existing scholarship reveals little about the security considerations of Anti-ELAB protesters. Ting~\cite[p.363]{SMS:Ting20} notes that networked protesters used ``encrypted messaging app Telegram and mass Airdrops over Bluetooth'' to coordinate protest activities, and that WhatsApp and Signal were used to share protest information and to request supplies. Ku~\cite{SP:SKu20} points to the mobilisation of Hong Kong youth activists through Telegram and the Reddit-like forum LIHKG, while Kow et al.~\cite{CHI:KowNarChe20} show how ``hundreds of groups'' on these two platforms were used to mobilise the protests through polls and the ability to act anonymously.
Importantly, however, none of these studies engaged with protesters, but relied solely on interpretative analyses of social media posts, forum posts and/or wider discourses.

\subsection{Higher-risk users and secure communication}\label{sec:high-risk-users}

Looking beyond large-scale protests, our research ties in with other qualitative works exploring the security concerns of higher-risk users. The use of secure messaging by higher-risk users is considered in~\cite{ErmHalMus17,HalErmMus18}. Through interviews with human rights activists and secure messaging application developers, this work outlines common and diverging privacy and security concerns among these groups. They found that while developers aim to cater to higher-risk users, the (perceived) security needs of these groups of users are not well understood and thus not well served. Similarly, in~\cite{AGRS15} the authors discuss the divide between activists and technologists. They advocate that ``security engineers [\ldots] step into the language of collective action within a political project'' to produce solutions that cater to the decidedly collective needs of activists and contrast this with a prevalent practice where ``in the absence of far away users under threat, designers can invoke them at will and imagine their needs''~\cite{AGRS15}.

The security needs of marginalised groups have received renewed attention from information security academics due to an invited talk by Seny Kamara at CRYPTO 2020~\cite{C:Kamara20,press:wired}. In this talk, Kamara characterises ``Crypto for the People'' as ``concerned with fighting oppression \& violence from Law Enforcement (Police, FBI, ICE), from social hierarchies and norms, from domestic terrorists''~\cite{C:Kamara20} and contrasts it with a libertarian-inspired concern for personal freedoms. More broadly, studies have explored security for civil society groups~\cite{SP:Scott16}, the security and privacy needs of journalists~\cite{LerZenRoe17,USENIX:MCHR15,PETS:McgRoeCai16}, privacy concerns among transgender people~\cite{CHI:LHKZSH20}, protection practices by Sudanese activists~\cite{SP:DSKB21}, fundamental security challenges experienced by refugees~\cite{CHI:ColJen19,CHI:ColJenTal18,CHI:JenColTal20,SP:SLIRK18} as well as undocumented migrants~\cite{CHI:GMSMTS18}. 
Like many of these prior works, our work suggests that the population we study has distinct (information) security needs that must be understood in order to design security technologies that meet those needs. 

\section{Preliminaries on technologies}\label{sec:preliminaries}

\textbf{LIHKG} is a Reddit-like forum that allows posts only from users with email addresses originating in Hong Kong (cf.~\cite{SP:SKu20}). \textbf{Signal} and \textbf{WhatsApp} are messaging applications that use phone numbers as contact handles and perform end-to-end encryption by default on all chats. Both applications support one-to-one chats as well as private group chats of up to 1,000 and 256 users respectively. \textbf{Telegram} is a messaging application that offers the option of end-to-end encryption for one-to-one chats only and supports public and private groups of size up to 200,000 as well as public channels with an unlimited number of subscribers. Telegram requires a phone number for registration but allows this to be hidden from other users.
\textbf{Facebook Messenger} is a chat service connected to Facebook, offering optional end-to-end encryption. On the technology level, Telegram makes roughly the same security promises as Facebook Messenger with respect to confidentiality -- with its bespoke MTProto protocol taking the role that TLS plays for Facebook -- but it makes it easier to adopt a pseudonym. 

Signal and Telegram secret chats allow users to send \emph{disappearing} messages which are deleted by the sending and receiving application after a certain time has passed (five seconds to one week). WhatsApp has recently enabled this option but has a fixed timer of one week. Telegram also supports \emph{scheduled} messages to be sent at a later date and time, before which the sending of the message can be cancelled.\footnote{The messages are scheduled on the server and thus will be sent even if the user goes offline afterwards.} Further, Telegram allows a user in a one-to-one chat to \emph{delete} messages for the other party, and a group administrator to delete messages for all group members. Neither WhatsApp nor Signal used to support this feature.\footnote{As of January 2021, Signal includes limited support for message deletion for everyone (only the sender can delete their own messages, within three hours of sending)~\cite{signal:deletion}, but this was not the case when the interviews were conducted. WhatsApp now supports the same feature with a time limit of one hour.} Telegram supports conducting anonymous \emph{polls} in groups and channels.

\textbf{Life360} is an application that allows remote monitoring of a phone -- e.g.~location, remaining battery -- that describes itself as a \emph{``family safety service''}~\cite{Life360} but is mostly known for being invasive~\cite{press:life360:wp}. WhatsApp and Telegram also support \emph{live location sharing} with another user for a period of time.

\section{Methodology}\label{sec:methods}
In this section, we outline our methodology, which is based on a qualitative research design and a grounded approach~\cite{Charmaz14,QHR:HenKaiMar17}, informed by existing social movement research (see e.g.~\cite{MSMR:BleTay02}). 

\subsection{Semi-structured interviews}\label{sec:methods-interviews}
Semi-structured interviews were chosen due to their exploratory nature; they are sufficiently structured to provide consistency across interviews and to address particular research questions, while leaving space for participants to offer new meaning to the topics (see e.g.~\cite{Galletta13}).

\subsubsection*{Interview process}
Informed by a topic guide\fullversion{ (Appendix~\ref{sec:interview-topic-guide})}{}, the interviews explored the use of communication technology within the protest environment and how protesters' security needs and practices shaped this use. Each interview covered topics such as communication technology use in Hong Kong, including specific platforms and applications as well as security concerns related to this technology use. The first two topics covered in the interviews deliberately did not focus on security, as it was important not to `force' a security angle. However, all participants mentioned specific security concerns related to their use of technology before we asked about them. This is not surprising, since information provided to participants prior to the interviews included information about the broader research focus and the composition of the research team\anonymous{(see Appendix~\ref{sec:pis})}{}. Moreover, the adversarial context foregrounded security concerns. Interview questions were intentionally broad to ensure that the research remained exploratory. This is an essential aspect of qualitative research, which works in the context of discovery and therefore emphasises openness and depth. The interviews were conducted by one member of the research team, between December 2019 and July 2020, as outlined in Table~\ref{tab:interviews}. Interviews were conducted remotely in English.

\subsubsection*{Participants and recruitment} 
11 participants from Hong Kong (P0-P10), all of whom had either primary or secondary experience of the protests, were recruited. All participants had attended at least one Anti-ELAB protest and were all members of protest-related online groups. The distinction between `primary' and `secondary' denotes front-line protest experience. Participants self-reported as `only' having secondary experience, because they had not been on the front line of a protest and were therefore less likely to have direct confrontation with law enforcement, while participants with primary experience had.

\begin{table}[h!]
  \footnotesize
  \caption{Participants \& Interviews}\label{tab:interviews}
  \begin{tabular}{ll @{\hskip 2em} lll}
    \multicolumn{2}{c}{\textbf{Participants}} & \multicolumn{3}{c}{\textbf{Interview}}\\
    \emph{ID} & \emph{Experience} & \emph{Duration} & \emph{Medium} & \emph{Timing} \\
    \midrule
    P0  & Primary   & 82 minutes & Audio & December 2019 \\
    P1  & Primary   & 43 minutes & Audio & December 2019 \\
    P2  & Primary   & 64 minutes & Audio & February 2020 \\
    P3  & Primary   & 51 minutes & Video & April 2020  \\
    P4  & Secondary & 47 minutes & Audio & April 2020 \\
    P5  & Secondary & 39 minutes & Video & June 2020 \\
    P6  & Secondary & 62 minutes & Video & June 2020 \\
    P7  & Primary   & 73 minutes & Audio & June 2020 \\
    P8  & Secondary & 53 minutes & Video & June 2020 \\
    P9  & Primary   & 87 minutes & Audio & June 2020 \\
    P10 & Primary   & 46 minutes & Audio & July 2020 \\
  \end{tabular}
  
  \vspace{1em}
We categorise participants' protest experience as \emph{primary} or \emph{secondary}, with the former defined as having been on the protest `front line'.
\end{table}

The protection of participants was our priority at all stages of the research. Initially, we only contacted publicly-known figures in Hong Kong, which led to three initial interviews. We then reached out to potential participants through two local gatekeepers,\footnote{See e.g.~\cite[Ch.3]{HamAtk07} for a discussion on the use of gatekeepers for access.} who shared our contact details and a participant information sheet (PIS) with potential participants\anonymous{(Appendix~\ref{sec:pis})}{}. The PIS outlined what participation would involve and how we would protect participant information. Gatekeepers were not involved in our communication with participants and whether someone decided to participate was not shared with them.

No specific selection or exclusion criteria were used to target individuals except for their primary or secondary involvement in the Anti-ELAB protests. However, this was by no means a straightforward recruitment process. We contacted more than 60 individuals linked to the protests and recruited 11. There are a number of reasons for this. First, the sensitive nature of the research and the importance of anonymity for protesters made it difficult to identify and recruit individuals with relevant protest experience. Second, parts of the research coincided with China passing a new national security law for Hong Kong, which also imposes restrictions on engaging with ``external elements''~\cite{press:ELAB:scmp}. Thus, many of our contacts declined to participate for safety reasons. Third, COVID-19 meant that travel to Hong Kong to engage with protesters was not an option. Hence, all engagements were carried out online. 

\subsubsection*{Human subjects and ethics}
All of our activities were approved for self-certification through our institution's Research Ethics Committee before the start of the research. Given the high-risk environment, and since our priority was to protect participants, we made sure to design our study in a way that minimised the collection of personally identifiable information. We recommended encrypted and ephemeral modes of communication, but followed participants' preferences, while using burner devices and anonymous accounts on our end to limit potential attack surfaces. Interviews were carried out by one researcher and were not audio recorded. With explicit consent from participants, extensive interview notes -- verbatim where possible -- were captured by the researcher. These were transcribed and stored on an encrypted hard drive.\footnote{Transcripts are retained for one year after publication and then destroyed.} To minimise risks to participants and researchers, we compartmentalised internally and only the researcher who carried out and transcribed the interviews has access to the raw data. Participants were not required to make their names known to us and we did not record any personal details in our interview notes. We do not report demographic information such as age or gender, nor do we report participant locations or their employment status. This is to protect their anonymity. Finally, participants were not compensated for taking part.

\subsection{Data analysis}\label{sec:analysis}
Interviews were analysed through an inductive analytical process, where the same (one) researcher coded the data through three coding cycles using NVivo 12~\cite{NVivo}. The first cycle used open coding and produced a range of descriptive codes, which were grouped in the second cycle to produce axial codes~\cite{Saldana15}. In the third coding cycle, the core variables in the data were identified and selective codes were produced and grouped into categories~\cite{CAQDA:RicRic95}. \anonymous{We provide a high-level example of the coding process in Appendix~\ref{sec:coding-tables}. }{}This form of analysis is employed to identify and analyse patterns across a qualitative data set, rather than within a particular data item, such as an individual interview. At the final stage of the analysis, technological implications were explored by the entire research team.

\subsubsection*{Limitations}

A number of limitations should be taken into account when interpreting our findings. First, our study was limited by the difficulties we experienced in engaging participants in our research, as outlined in Section~\ref{sec:methods-interviews}, and research findings might have captured other practices if further interviews had been conducted. Yet, the semi-structured nature of the interviews was chosen to provide depth rather than scale. Moreover, the analysis suggests that coding saturation was reached. Second, conducting interviews online limited the researcher's ability to observe the participants' physical settings, which might have affected their ability to speak freely. Third, some protesters, who declined to participate, might have been particularly concerned about security. Fourth, while participants spoke fluent English, it might have been possible to recruit a broader selection of participants if interviews had been conducted with the assistance of a translator.

Finally, there is an inherent bias in interview-based research, particularly when it concerns security or technology questions, given that participants self-select to take part. Some contacts decided against participation because they did not feel that they knew enough about the technologies they were using. This limitation is not unique to this study, but mirrors other technology-focused interview-based studies; they are inherently biased towards the more tech-savvy end of the population being studied, such as security trainers or attendees of IT security trainings. Future work should consider adopting ethnographic methods of inquiry to overcome this limitation.

\section{Research findings}\label{sec:findings} 

Our research findings are structured into five subsections: Section~\ref{sec:tools} focuses on the technologies used by protesters and why, Section~\ref{sec:social-organisation} shows how these technologies interact with the social organisation of the protests, Section~\ref{sec:forward-secrecy} discusses tactics for detecting and reacting to arrests, Section~\ref{sec:limitations} shows how protesters address the limitations of the technologies they rely on, and Section~\ref{sec:ideas-security} focuses on how and from where protesters develop ideas about their security.

\subsection{Tools}\label{sec:tools}

Internal communication between Anti-ELAB protesters was mainly done through two messaging applications: Telegram (predominantly) and WhatsApp, with most protesters joining dedicated protest-related groups on both applications.

\emph{Telegram} was used by all participants and dominated our findings. One participant summarised Telegram as \emph{``the most useful platform, followed by WhatsApp''} (P0), while another expanded: \emph{``For communication and organisation, most people use Telegram''} (P6). Participants observed that its popularity in the protests was based on three conditions: (1) its widespread adoption prior to the protests, (2) its security, which was perceived to be better than any other messaging application and (3) the ability to form both large and small groups. Telegram's polling feature
emerged as another reason for adoption as well as various of its features used to monitor fellow protesters for arrest, as discussed in Section~\ref{sec:forward-secrecy}. Participants understood Telegram to give them the \emph{``most security''} in group chats (P0). As explained by one participant: \emph{``We have a group on WhatsApp and another one on Telegram, but we use the one on Telegram to talk about our actions [\dots{}], because we think Telegram is more secure''} (P9). One participant (P5) noted that, although end-to-end encryption was not the default setting in Telegram group chats, this could be enabled. This is incorrect (see Section~\ref{sec:misconceptions}) and demonstrates how an incomplete or, as in this example, incorrect understanding of security might shape participant perceptions.

\emph{WhatsApp} was also used by the majority of participants in our study and they assumed that this would be the case for others too: \emph{``most protesters use WhatsApp too, yes definitely''} (P3). Yet, WhatsApp was seen to be less suitable compared to Telegram because it only allows for groups of up to 256 members.

While \emph{Signal} was brought up by several participants without prompting, our data suggests that it has not seen any significant adoption among Hong Kong protesters. Participants highlighted the discrepancy between what they perceived as their security needs and what is offered by Signal. First, the need to provide a phone number was seen to conflict with the need for anonymity to avoid police detection: \emph{``the reason we don't use Signal is because Signal requires that you know the telephone number of the other people if you want to make a contact''} (P7) and \emph{``The thing is, people in Hong Kong cover their faces when they go out to protest. They want to be anonymous. So, if you have to then give your phone number, it doesn't make sense''} (P7).\footnote{Anti-ELAB protesters defied the ban on wearing face masks that was introduced in Hong Kong in October 2019~\cite{press:HK:NYT-2}.} When asked whether they would consider using burner SIM cards to use Signal, they responded that the benefits would not outweigh the risks. Second, the function of being able to delete messages sent by other group members was key for protesters: \emph{``You cannot tell people to use Signal instead of Telegram, because that's not realistic and also Signal is horrible at other things that the protesters need. For example, you cannot control what happens to your messages once you have sent them. You can just use disappearing messages''} (P6). Thus, participants in our study compared the security offered by Signal to Telegram -- not to WhatsApp -- when making decisions about which tools to use.

While WhatsApp also requires phone numbers, it was already widely used by participants before the protests and they felt confident and, as a result, secure using a tool with which they were already familiar. Where Telegram catered to their need for anonymity in large group chats, WhatsApp was used for small close-knit groups, where anonymity was not a security need. Hence, Signal was not seen to provide them with additional security or required key functionality. 

\subsection{Social organisation}\label{sec:social-organisation}
Our work speaks to the utility of groups on messaging applications for on-the-ground protest organisation enabling collective practices, strategies and tactics -- and to related security requirements. Here, we discuss such practices and show how different types of groups, characterised by their size, imply different, at times opposing, security requirements.

\subsubsection{Group types}\label{sec:group-divides}
Two types of groups were identified in the data: large Telegram groups, sometimes with 2,000, 20,000 and 50,000 members and small(er) groups on both Telegram and WhatsApp. The former comprised public groups set up to disseminate protest information across large networks, facilitate collective decision making and reach and connect disparate groups. The latter were formed around more or less close-knit groups of protesters.

All participants in our study were members of several Telegram groups; some small groups, made up of people they had met during the protests, and some large groups, which they predominantly used for information-gathering purposes. This divide also mirrors the division between participants' protest experience; those with only secondary experience had never been part of small protest groups, but were in several large public Telegram groups. Participants with primary protest experience were members of both types of groups. All participants, regardless of protest experience, gave examples of how they knew that the large Telegram groups were infiltrated by e.g.~local police officers, who monitored the groups to gather information about protesters and protest strategies. Several participants also reported deliberate attempts to undermine the protest efforts in these groups by presumed infiltrators.
While there was general consensus among participants that the disruption caused by these infiltrators was minimal, it highlights an important aspect of big group chats: all participants accepted that confidentiality could not be achieved in these large groups, while they assumed that it could be achieved in the smaller groups. However, large groups were essential for the successful organisation of protest activities because of their scale and reach -- and crucial for the collective actions that they facilitated, such as joint decision making.

For all participants with primary protest experience, being able to organise quickly and securely was the key motivating factor behind having smaller rather than larger groups. The large groups were run by dedicated administrators (see Section~\ref{sec:group-administrators}), while the small groups were formed \emph{``quite organically and not that organised''} (P5). Each small group, however, had its own identity, its own utility. One participant explained this by drawing on two groups, one with 26 members on Telegram and another one with six members on WhatsApp: \emph{``there are still some differences between those 26, because I met six of them and formed a small team. But the other 20 joined later. So, actually, those 20, I haven't met them before, face-to-face. We have the WhatsApp group, only the six of us. And on Telegram we have the 26''} (P2).

\subsubsection{Strategies and tactics}\label{strategies-tactics}
The importance of secure messaging applications for protesters has already been articulated in previous works, e.g.~\cite{ErmHalMus17,GiArAl19,HalErmMus18,Trere20}. In the Anti-ELAB protests, such applications more specifically cater to the particular strategies and tactics employed by protesters: a flat structure, mobile, dynamic and large-scale in nature. All participants in our study explained how the ability to collectively decide on strategies and tactics in real time across large and geographically dispersed protest sites was essential to the success of the protests. One participant articulated how Telegram provided a \emph{``safe online space''} to collectively decide specific actions: \emph{``we use Telegram to talk about our actions, our equipment, our strategies, our tactics''} (P2). Another participant spoke about how Telegram enabled immediacy, which was needed when tactics had to be altered during a protest: \emph{``during the protests themselves, the information is more related to strategy, like, what to do right now''} (P5). Both quotes highlight the sense of urgency felt by participants when talking about sharing tactical information during protest actions.

Several other participants expressed a sense of information overload given the volume of information being shared during protests. This often made it difficult for them to keep up with evolving protest tactics. One participant noted: \emph{``When protests are actually taking place, the groups are much more active, there's information all the time and it's difficult [\dots] to know what the strategy is''} (P9).
Such statements exemplify the challenges experienced by protesters when faced with multi-directional and extensive information in both adversarial and highly digitalised environments: \emph{``it's hard to keep track of stuff''} (P10). All participants with primary protest experience spoke of how they would have to make tactical decisions within seconds when receiving information about police locations or new gathering points. For many, this meant deciding which groups to \emph{``keep open and which to close''} (P7) while participating in protest activities, hence, limiting the information they would have to digest.

\subsubsection{Collective decision making}\label{sec:decision-making}
Protests are by their very nature a collective endeavour and the mobilisation of protesters has been the topic of many recent works, as identified in Section~\ref{sec:large-scale-protests}. However, beyond mobilisation, our data reveals how Telegram and LIHKG were used to make collective decisions about protest tactics, in real time.

Several participants in our study exemplified how large Telegram groups were used to vote on \emph{``the next move''}, as explained by P7, while LIHKG was used to vote and decide on broader protest strategies at the start of the protests. \emph{``This forum called LIHKG\@. We used it for strategy and stuff. Like in Reddit, people can vote [\dots] And we used it because you can only register with a Hong Kong email provider''} (P9). These features -- collective and limited to people with a Hong Kong email account -- made LIHKG a central platform early on in the protests. One participant suggested that it enabled \emph{``nuanced discussions about strategy and to vote on strategy''}~(P5). Yet, many participants noted that, over time, the organisation of on-the-ground actions \emph{``couldn't be done on the forum because the police is monitoring it''} (P9). Thus, for real-time voting on tactical moves during protest actions, protesters had moved to Telegram groups, where polls on, for example, \emph{``where to go next''} (P10) often received several thousand votes. While all participants in our study also assumed police monitoring of the public Telegram groups, the speed with which collective decisions could be executed made police infiltration less of a concern. Forums were, on the other hand, generally seen to be slow and not suitable for live protest action.

One participant explained how the voting worked best when only a few options were given, enabling protesters to make a \emph{``simple choice between A or B''} (P3). However, based on our data, we see that the option with the most votes is rarely followed by everyone. Given the anonymous nature of these groups and of the polls -- and since anonymity was a key security need for Anti-ELAB protesters -- it is unclear who votes in these polls. The scale of these groups was, however, critical for the success of the protests for two main reasons: it established a strong sense of collective decision making which, in turn, meant that no single person was seen to be publicly leading the protests. For the protesters, this had a security function as well, as it was seen to spread the risk of arrest to several thousands of people; to everyone who voted.

\subsubsection{Group administrators}\label{sec:group-administrators}
The centrality of protest groups on messaging applications meant that group administrators occupied key positions in the protests. Without public leaders, our data suggests that group administrators were seen as the leaders of the protests. While not directly articulated by the participants in our study, many of them spoke to the multiple and critical roles performed by group administrators and the trust that protesters placed in them. Importantly, however, group administrators remained anonymous leaders, hiding their identity to avoid police detection. Moreover, most groups had several administrators to \emph{``spread the risk [for the group] to more than one person if one admin is compromised''} (P9), allowing non-compromised group administrators to revoke the administrator capabilities of those compromised. The same administrator also often managed several groups at the same time through different accounts.

Our data contains several examples that support the interpretation that administrators took the role of leaders. One participant noted: \emph{``We have groups for voluntary medical support, and we have many groups for legal support. So, the whole protest, without leaders, is organised by these group administrators''} (P9). This mirrors how many participants experienced the protests themselves: as a decentralised movement, with \emph{``many people who lead but no organisation''} (P3) or \emph{``flat but not leaderless''} (P2).

To illustrate the central role of administrators, we use an example that was recounted by all participants in our study: a voluntary ride-sharing scheme. This was critical to get protesters (``frontliners'' in particular) to/from protest sites, as using public transport was \emph{``too dangerous for protesters because the police go to public transport to attack and arrest people''} (P3). However, many participants noted that the scheme required protesters to trust the administrators of the groups through which the scheme was run and their vetting procedures, which relied on drivers sharing their licence details with the group administrator(s). This was a way for them \emph{``to verify the driver's identity before referring them to the protesters''} (P2). When a protester requested a driver through the group, the administrator would \emph{``link up the car/driver and me as a protester. We don't know the driver or the administrator, but we know the licence number''} (P7). Some participants noted that while administrators would try to verify the driver's identity before referring them to protesters, they knew of several examples of undercover police officers pretending to be drivers, resulting in arrests. Still, participants with primary protest experience had all used this scheme and said, in different ways, that they had \emph{no choice} but to trust.

\subsubsection{Onboarding practices}\label{sec:onboarding}
The practice of establishing close-knit groups on Telegram and WhatsApp led to a number of security constraints for protesters, which centred on the need to establish trust within highly digitalised and adversarial environments. All participants with primary protest experience noted how their groups had developed particular onboarding practices rooted in interactions at sites of protests. This was seen as necessary to verify the identity of any newcomer to the group and ensure trust among group members. Based on the experiences of the participants in our study, specific onboarding practices were adopted for both Telegram and WhatsApp groups with between five and 30 members.

Our data shows how small close-knit groups were formed around protesters who had met face-to-face during the protests \emph{``before moving the connection online''} (P4), as \emph{``seeing each other and standing on the front line together is very important for trust''} (P10). These trust bonds were described to be established through shared aspirations and were seen to be key for the success of the protests as they enabled affinity groups to form and carry out essential tasks, e.g.~provide legal or first aid. This was supported by another participant, who noted that it was important for their group that any new members supported their faction: \emph{``So we see them in person first and we then also know that they are chanting the right slogan''} (P9). Participants also explained how offline connections would only be moved online once rapport had been established with new group members. Our data suggests that, for most groups, this form of gradual onboarding to establish trust sometimes took weeks and sometimes months.

We unpack this collective process by using an example given by one participant, who belonged to two small affinity groups. They explained: \emph{``First, we have to meet them face-to-face. It's not that you just meet them and then add them, it's about values and beliefs and aspirations. We want those newcomers to work with us in the field several times first. If they share the same beliefs and aspirations, they can officially join our Telegram group''} (P0). For the close-knit groups, where specific protest activities related to the group would be discussed (what protesters deemed \emph{``sensitive information''}), all existing group members would have to meet any new group members before they would be allowed to join. 

Our data contains some examples of specific onboarding processes where some group members had been unable to meet a new group member. This would then become a negotiation between existing group members: \emph{``someone in the group will say `I know a person who might be able to contribute to this group', and there will then be a short discussion and then a decision''} (P3). Participants noted that while this was not \emph{``bullet proof''} (P10), it was also important for them -- and for the success of the protests -- to accept group members who they thought would be able to contribute to their efforts. However, this form of onboarding was accompanied by a level of distrust for some participants, who would insist on meeting all potential group members before accepting them into the group:  \emph{``I would want to meet all group members in person first, before accepting them''} (P1). As expressed by another participant: \emph{`` Sometimes you have to make a choice, even if you haven't got enough manpower, you only recruit people who you trust''} (P10). The main concern was articulated as \emph{``potential infiltration of police''} (P7). This was a common worry expressed by participants and was connected to their experiences with large Telegram groups, where police infiltration was explained to have led to several arrests.

\subsection{Indicators of compromise}\label{sec:forward-secrecy}

Our data demonstrates that the threat of arrest during a protest and the subsequent compromise of the arrestee's close-knit affinity group was a key concern for participants. Our data shows that different protest groups adopted subtly different approaches to monitoring each other while attending protests. Our data also suggests that this was a widely adopted collective (security) practice for Anti-ELAB protest groups.

Our data contains three approaches to monitoring: the use of specific monitoring applications, scheduled messages or regular messages. The use of specific live-tracking applications was practised by several participants and comprised a system whereby when some group members went onto the street, the rest of the group would be responsible for monitoring their whereabouts using WhatsApp or Life360. Some participants explained how they would use both applications simultaneously to ensure that they would be able to receive constant updates. This was seen as particularly useful to determine whether a group member had been arrested: \emph{``There are some signals that tell me that the person got arrested. For instance on the live location, if they disappear from the map then I know something is wrong [\dots] if I know they have battery and suddenly disappear then I can call them. If no-one picks up the phone for a long time and we can't find them in the field, then we will track their last location. And then we know whether they have been arrested''} (P1).

Another participant detailed their group's approach to live monitoring, which relied on regular messages: \emph{``If my friends go out in the protest, I'll stay up and every hour I'll text and ask `are you safe?' And if they don't respond within two-three hours I'll assume that they are arrested''} (P3). The same participant reported that \emph{``there's a feature in Telegram that allows you to periodically send out a message. So, it does something automatically periodically -- so these pings are exchanged among a group and if you see that someone isn't responding to the ping, then probably something bad has happened''} (P3).\footnote{This is not a feature included in Telegram as described, but note that bots~\cite{telegram:bots} may be used for this purpose as they allow to expand the functionality of the application when added to a chat.} Another group used timed or scheduled messages to alert group members should their phone be inactive for a period of time: \emph{``we use timed messages, so others know that if they receive the message, I'm probably arrested''} (P9). That is, protesters would schedule a message to be sent later and would cancel this scheduled message once they returned from the site of protest. If they failed to cancel the message, this was taken as an indication of a problem.

Other participants gave similar accounts and noted that these practices had been systematised within many groups -- and that groups had learned from each other -- in response to a growing number of arrests. For them, being able to monitor each other was seen as a way \emph{``to protect others when someone gets arrested and also to provide legal assistance''}~(P3). For all participants in our study, this form of monitoring was important to protect and support group members in the event of arrest: first, by arranging for legal aid and, second, to control access to information about or related to other group members. It is for this reason that the ability to delete messages sent by any member in a group was seen as vital. In case of an arrest, the group administrator(s) were responsible for removing messages from the arrestee's device and to remove them from the group. This feature was seen as key: \emph{``I can delete the messages for others, not only for myself''} (P7); as allowing them to \emph{``control the conversation''} (P4) or to \emph{``control what happens to your messages''} (P5) and to kick out anyone who had been arrested and to delete all group messages -- \emph{``so we can at least keep the others safe''} (P2).

Our data highlights a number of concerns and conflicts raised by participants in relation to such live monitoring practices. First, the concern that their live locations might become available to the police showing that they had \emph{``committed crimes by being in locations they aren't meant to be''} (P7). Thus, this appropriation of consumer applications with unclear privacy guarantees illustrates the limitations of existing security technologies. Second, live monitoring through specific location-tracking applications was also seen to limit participants' control over access to data as it is not possible to delete the data in Life360 or WhatsApp: \emph{``if a group member is arrested, the police can track the others via the app as we cannot delete for others''} (P2). More broadly, participants articulated how they would try out different technologies to find \emph{``the best solution available''} (P5), but also know that these did not serve their security needs. We expand on this point in Section~\ref{sec:ideas-security}.

\subsection{Limitations of technology}\label{sec:limitations}

We present the additional practices adopted by protesters to address the limitations of the technologies they use. Protesters spread their identities across different accounts and devices to achieve a level of pseudonymity and a variety of low-tech tactics were adopted to handle congested networks. 

\subsubsection{Pseudonymity}\label{sec:protecting-pseudonymity}

All participants in our study spoke about how their involvement in the protests had heightened their focus on personal and information protection. For participants, particularly those with primary protest experience, any personal information was considered sensitive.
In security terms, their (online) identity was closely tied to their protest activities, driving a growing need for pseudonymity: \emph{``protesters make their profiles private, they use a separate SIM card, they use pseudonyms and so on''} (P6). Several participants explained how protesters had \emph{``a separate phone when [they] go out and a separate SIM card''} (P4) and how they had \emph{``another group with a different number which is attached to a different SIM card and completely isolated from the usual groups''} (P2). This separation between protest groups and phone numbers was seen as a key mechanism for protecting individual anonymity and to go undetected by the police: \emph{``So, that's why we don't want to give out phone numbers, even with burner phones''} (P9). Another participant articulated how they, along with other group members, had several phones and other devices as well as several accounts on different applications. This is in addition to several protesters sharing one account, which was said to be done to ensure that others \emph{``won't know they are not the same person''} (P10).

These desires to protect their identity and the identity of group members, combined with what many participants referred to as increasing surveillance measures by Hong Kong authorities, were articulated as causing a critical need for anonymity. This need was also linked to the popularity of Telegram as a protest tool in Hong Kong: \emph{``I think Telegram is particularly good because it allows you to stay anonymous''} (P5). Yet, participants also noted how the \emph{``move to Telegram''} had created a \emph{``conflict between trust and anonymity''} (P9) because they were no longer able to \emph{``look at people's Facebook profiles''} (P7) to establish their identity; a practice that was used extensively during the 2014 Umbrella Movement. Hence, online vetting of potential group members had become impossible.

\subsubsection{Disconnected discontent}\label{sec:offline-needs}

All participants with primary protest experience had also experienced being disconnected, due to network congestion, while taking part in protest activities. They explained how they had found alternative ways of communicating with other protesters. These took different forms.

First, some participants with primary protest experience articulated how they relied on interactions with other protesters in the street, which enabled them to develop and use hand signals to pass on messages: \emph{``Sometimes it's just much easier just to wave or communicate using some hand gestures, when the network is down''} (P10). Participants gave specific examples of this form of non-verbal communication. They noted that hand signals were often used to communicate which supplies were needed on the front line: \emph{``If you see someone doing a cutting motion with these two fingers [index and middle fingers] you know that scissors are needed''} (P9). Arms orbiting the head was said to indicate that helmets were needed on the front line (P7). Second, some participants spoke about how they would go to places with WiFi facilities to try to send messages during the protests. Yet, this approach was only adopted at critical points when they saw no other ways of communicating. Third, some participants noted how they would \emph{``revert''} to using SMS, at times when they could not connect to the Internet. Exemplified here by one participant: \emph{``there was a time when I was at [location] because of the protests and couldn't connect to a network for some reason and couldn't connect via Telegram or WhatsApp. So, we could only connect with the outside via SMS\@. Paid messages''} (P2).

Finally, most participants had heard about the mesh networking application Bridgefy (see~\cite{EPRINT:ABJM21}), which according to news reports saw a spike in downloads in Hong Kong in September 2019. However, none had successfully used it: \emph{``it just doesn't work''} (P7).

These alternative approaches of connecting when the Internet is not available speaks to the disconnected needs of Anti-ELAB protesters. While Hong Kong authorities did not resort to shutting down the Internet, protesters experienced significant disruptions to their digital communications. These disruptions, which are a feature of the protests' large-scale nature -- \emph{``A million people just makes it impossible to communicate''} (P9) -- render the technologies that protesters rely upon largely futile, at the height of protests.

\subsection{Routes of security perceptions}\label{sec:ideas-security}

We explore where Anti-ELAB protesters' notions and ideas about security and their own security needs have come from. In so doing, we first show how previous protest experience shapes protesters' practice of security and how the adoption of messaging applications is a result of a change in security mindset among protesters. Second, we show how protesters with no or limited protest experience adopt the technologies and practices employed by more experienced protesters. It is worth noting, however, that our data reveals that participants with only secondary experience of the protests assumed greater adoption of applications such as Bridgefy and Signal than what was exemplified by participants with primary protest experience. This is not surprising given how (inter)national media outlets have reported on some of these technologies~\cite{press:HK:forbes}. Yet, it is important to distinguish between actual and perceived adoption and requirements, and it points to the urgent need for secure technology designers to engage with the groups of users they seek to serve, as also noted in~\cite{AGRS15}.

\subsubsection{A shift in security mindset}\label{sec:security-mindset}
Our data suggests a change in protesters' security mindset during the Anti-ELAB protests, with most participants highlighting a growing need for anonymity, due to heightened surveillance, and confidentiality, in relation to trusted and close-knit small groups. All but one participant with primary protest experience had also taken part in previous protests in Hong Kong and had experience of using technology within such protest environments. These participants compared their experiences in the current protests with those of the 2014 Umbrella Movement, where \emph{``you basically had no access to the Internet as there was so much traffic and the network was super slow''} (P3) and \emph{``most was organised over Facebook''} (P2). In addition to changes to technology, several participants highlighted how the protest environment had become increasingly adversarial: \emph{``In the 2014 movement, things happened much more slowly [\dots{}] There was no conflict most of the time. But this is very different now''} (P9). Many participants noted that this had led to a shift in security mindset among protesters. While \emph{``before June last year [2019], people would be gathering on Facebook''} (P6), \emph{``just talk about about sensitive information on Facebook's messenger''} (P10) and \emph{``not think about end-to-end encryption''} (P2), this had changed with what they described as an increase in police surveillance and arrests. This shift in mindset had led to a greater adoption of Telegram.

\subsubsection{Collective information security}\label{sec:collective-needs}
For Anti-ELAB protesters, as articulated by the participants in our study, information security is a collective endeavour. It is practised by individual protesters, who have their own security perceptions and needs, yet these are shaped by the security decisions of the group. At a high level, this is not surprising given the centrality of groups in these protests, the practice of voting on strategies and tactics, and the fact that not everyone holds the same security knowledge. It does, however, speak to how security is practised within groups.

It also demonstrates that, to be a group member, protesters have to buy into the security collectively decided for the group. One participant explained how they had tried to convince members of their group to switch to Signal after they had realised that \emph{``people in other countries use Signal''} (P2). Yet, this had been unsuccessful as other group members preferred to keep the group on WhatsApp, as they were already familiar with this application and its (perceived) security. This led to them having to compromise their own security needs to be a group member. One participant said that they had changed their practices to be in line with other group members: \emph{``I only started to use Telegram during these protests. I didn't use it before. I heard that Telegram is used by terrorists, because it is so secure. And it is used by my groups''} (P1). This participant accepted that they \emph{``had to conform to be in the group''}. Participants explained how they had observed others \emph{``change their security mindset''} to buy into the security of their group~(P3).

Our data also contains several examples of how participants were either unsure about the level of protection offered by some of the technologies they used or knew that a particular application was not \emph{``the most secure''} (P10). For example, one participant explained how they had accepted that they could not \emph{``do everything to protect''} themselves (P9). This was reiterated by another participant: \emph{``I do not know if Telegram or WhatsApp are safe to use or whether the Chinese government can listen in, but I use them because others use them''} (P7). Moreover, some participants had accepted that their security needs would not be met by the technologies they used but that they offered \emph{``good enough''} security (P0).

Participants with less protest experience or who did not perceive themselves to be security conscious noted how they relied on other protesters for advice. At a group level, the security approaches and technologies adopted by one group would often be adopted by another group. This is evident from comments made by participants about how they would look to more established groups for security advice. Our observations about onboarding practices and live location monitoring also exemplify this point. First, onboarding processes adopted by groups were generally performed in similar ways. Second, live location monitoring was practised by all groups that included participants with primary protest experience. These subtly different approaches centred on only a few technological solutions and established practices.

\section{Discussion}\label{sec:discussion}

In this section, we reflect back our findings to information security scholarship, with a focus on cryptography.

\subsection{Secure messaging}\label{sec:target-selection}

\subsubsection*{Telegram} The participants in our study reported Telegram as the predominant messaging application used by Anti-ELAB protesters. This finding is corroborated by media reports, e.g.~\cite{press:HK:telegram:bloomberg}, and corroborates prior work that established the use of Telegram by activists~\cite{ErmHalMus17}. However, Telegram has received relatively little attention from the cryptographic community~\cite{JakOrl16,PhD:Kobeissi18} or information security research~\cite{AKPKKMTS17,AngCanGua17,SusKok17}. As noted in~\cite{PhD:Kobeissi18}, academic attention is focused on the Signal Protocol partly due to its strong security promises such as forward secrecy and post-compromise security. Indeed, even when Telegram is studied, its end-to-end encryption in secret chats is the focus, cf.~\cite{JakOrl16,PhD:Kobeissi18}. This feature, however, has little impact on the actual security provided by Telegram in the use case considered here, since secret chats are one-to-one only.
Group chats are secured at the transport layer by Telegram's bespoke but understudied MTProto protocol, which Telegram typically uses in place of TLS\@.\footnote{In~\cite{ErmHalMus17} it is incorrectly reported that group chats default to TLS.}
Telegram also implements a variety of features meant to support anonymity within groups, often in response to user demand~\cite{tlg:anon-num, tlg:anon-admin}, which have not been rigorously examined. Our work suggests the study of MTProto and the anonymity guarantees of Telegram's group chats as pressing problems for future work.

\subsubsection*{Messaging Layer Security (MLS)} Our findings support the decision by the MLS working group to support groups of up to 50,000 users~\cite{STKCO18}. On the other hand, our findings indicate diverging security goals for different types of groups, roughly characterised by their size, in the setting under consideration: anonymity of group members towards each other but no confidentiality in large groups forming one type, and another one being confidentiality and authentication in small, close-knit groups. Our data presents a use case where a hierarchy of permissions in groups is central and where out-of-band authentication of group members may be assumed, weakening the need to trust the Authentication Service as defined in~\cite{STKCO18}. MLS does not model group permissions at a cryptographic level but aims to be compatible with this use case when such restrictions are externally enforced. It is worth noting that MLS supports multiple devices per user, while our data presents the practice of multiple users sharing the same account. It is plausible, though, that this conceptual difference does not make a difference in practice on the MLS level.

\subsection{Security notions}\label{sec:security-notions}

\subsubsection*{Compromise} In the literature, the notion of \emph{forward secrecy} (FS)~\cite{EC:Gunther89a,C:Krawczyk05} is understood as the protection of past messages in the event of a later compromise of an involved party and the notion of \emph{post-compromise security} (PCS)~\cite{CSF:Cohn-GordonCG16,EPRINT:CreHalKoh19} as the protection of future messages some time after a (usually full state) compromise. Both of these security notions work with a persistent, global adversary of some form. Post-compromise security protects against an (ordinarily at some point passive) adversary after a compromise. Forward secrecy protects against an adversary that either passively observed the communication (weak FS) or even actively attacked it before the compromise.\footnote{Social dimensions of targeted attacks (active) and mass surveillance (passive) are discussed e.g.~in~\cite{GueKunVan16,JagSyv17,C:Kamara20}.}

The compromise the participants in our study were most concerned about was during and after an arrest. Here, they were concerned with both forward secrecy (remote message deletion) and post-compromise security (excluding an arrestee from a group). However, their notions differed from those in the literature. First, a cryptographic scheme achieving forward secrecy would not achieve the notion of forward secrecy desired by the participants in our study as messages remained stored on the recipient's device.\footnote{Disappearing messages only provide a partial solution, leaving messages received within the expiration window exposed.} That is, our participants assumed and aimed to protect against a compromise that reveals not only key material but also the entire chat history (stored on the phone). Second, a security goal of the participants in our study was to protect themselves during the compromise not just afterwards. As indicated in our research findings, there is a variety of behaviours attempting to detect and control compromise \emph{as it happens}, including location monitoring, timed messages, revocation of administrator capabilities and message deletion for others, all done on behalf of the compromised person by the remaining group members (we discuss the resilience of these methods in Section~\ref{sec:misconceptions}). Critically, their notion of post-compromise security was at a group level (removing the compromised party) rather than for the compromised party.\footnote{It is worth noting that the grounding of authentication in offline interactions and the assumed detectability of a compromise provides a mechanism to achieve some form of post-compromise security in the more traditional sense out-of-band, possibly at the cost of replacing a burner phone and/or chat group.}

Overall, the adversary model of the participants in our study is both stronger (the adversary also compromises the chat history; protection against an adversary during a compromise is intended) and weaker (detectable) than those in the literature, i.e.~the resulting security notions are incomparable.

\subsubsection*{Time and place}\label{sec:time-place} Implicit in our data is that security and access requirements change with time and place. Group members away from the front line are assumed to be relatively safe, compared to those on the front line facing immediate arrest. This suggests a partial solution for forward secrecy. Group membership could be restricted while out in the field -- e.g.~messages disappear faster, no access to the list of group members, only pseudonymous handles, no admin rights -- with fuller access being restored using a secret-shared key afterwards.\footnote{This would partially mirror the practice adopted by some business travellers to move their data across borders online to avoid confiscation at the border, the latter being a use case used to motivate PCS in~\cite{EPRINT:CreHalKoh19}.} More broadly, it suggests modelling the dynamic nature of access privileges over time and place.

\subsubsection*{Anonymity and authentication} The use of forums such as LIHKG and large public Telegram groups, combined with the desire to avoid being tracked, suggests a need for a different kind of communication platform. If infiltration is assumed, the focus shifts from protecting confidentiality to protecting identity.
As our data shows, this focus on anonymity surfaces the question of how to establish trust. A number of proposals exist in the literature: Dissent~\cite{CCS:CorFor10} claims a ``collective'' approach to anonymous group messaging with accountability, Riposte~\cite{SP:CorBonMaz15} aims to provide a secure whistleblowing or microblogging platform that resists disruption and AnonRep~\cite{NSDI:ZhaiWCSTF16} presents an anonymous reputation system for message boards. The systems vary in cryptographic assumptions, threat models as well as ability to scale, but none of them provide real-time messaging and are hence only suitable for public forums that are not time-sensitive. None of the cited works have moved beyond the prototype stage, and many open research questions remain in the area.

Closely related is the study of reputation systems, whether centralised~\cite{FC:BloJuhKol15, AFRICACRYPT:GarQua19} or decentralised~\cite{iTrust:PavlovRT04, FGCS:AzadBH18}, originally motivated by the information leakage in services such as eBay or Uber which utilise public user ratings. It is not immediately clear how such a system could be translated to the setting of user trustworthiness in anonymous messaging, but the emergence of crowdsourced services such as the voluntary car scheme reveals potentially more straightforward applications. Yet, the context in which reputation systems are reasoned about is largely limited to marketplaces and cryptocurrencies.
Moreover, given the strong emphasis on collective or group action indicated by our data, it is an interesting open question where (if anywhere) group~\cite{EC:ChaVan91} or ring~\cite{AC:RivShaTau01} signatures, the primitives often underlying reputation schemes, may productively be deployed. However, the high level of mutual trust required to operate in small affinity groups and the practice of sharing account credentials might make the functionalities of these primitives unnecessary.

\subsubsection*{Trusted third parties}
Our data indicates that the Anti-ELAB protests rely heavily on trusted third parties. This is true in a technological sense, e.g.~group chats are not end-to-end encrypted and facilitated by Telegram's servers, which are protected by geopolitics, i.e.~the limited reach of the current adversary. This observation corroborates prior work on activists~\cite{ErmHalMus17}.

While this technological reliance might be an artefact of necessity -- viable alternatives are absent -- our data also shows that trusted third parties, in the form of anonymous group administrators, are a central feature of these `decentralised' and `leaderless' protests. The work of Azer et al.~\cite{NMS:AzeHarZhe19} highlights the significance of what they call ``connective leadership'' in digitally enabled and self-organised contemporary activism. Echoing this work, our findings illustrate how even `leaderless' protests require leaders to connect protesters and protest groups. In the Anti-ELAB protests, due to their highly digitalised nature and experiences from the 2014 protests, group administrators act as connective leaders. This makes understanding their information security practices and needs a critical area of research for information security researchers, as the compromise of one of these administrators can have significant consequences, see e.g.~\cite{press:hk:stand}. This is particularly pertinent as large-scale protests around the globe adopt the strategies developed in these protests -- their dynamic, mobile, digital and flat structure. On a technological level, recalling that the administration duties are often split between different individuals, and that the most prevalent form of compromise -- arrest -- may be detectable, MPC solutions, even in the efficient non-malicious setting, might suggest themselves.

\subsection{Misconceptions}\label{sec:misconceptions}
The participants in our study made security decisions based on specific functionality needs and explicitly formulated domain-specific security perceptions. However, our data reveals several mistakes in their perceptions of the security guarantees of the tools they relied on. Participants assumed that end-to-end encryption could be enabled in Telegram group chats, which is incorrect. The data also highlights that the ability to delete messages on other users' devices and to remove them from a group after an arrest drove the adoption of messaging platforms. Yet, these tactics assume that the compromised device continues to receive and process deletion requests; the more this tactic catches on and thus registers with the adversary, the more dubious this assumption becomes. Such misconceptions are not unique to our study. For example, several studies on usability, e.g.~\cite{SOUPS:Ion15,USEC:Vaziripour18}, highlight user misconceptions and false mental models in relation to security. Other studies, e.g.~\cite{SP:ASBDNS17,EUROSP:DNDS19}, also suggest that users find it difficult to understand the security of the applications they rely on and whether it fulfils their needs. For higher-risk users such misconceptions can have dire consequences for their safety, especially since the misconceptions identified in our study tended to overestimate the security guarantees given. Critically, however, our data highlights the negotiated and collective nature of adoption in this setting, in contrast to individual preferences foregrounded in previous work.

\subsection{Collective security}\label{sec:collective-security}

Our findings speak to an understanding of information security that rests on collective practices, where security for the group is negotiated between group members and where individual security notions are shaped by those of the group. They show how Anti-ELAB protesters practised security to fulfil their own security needs as well as those of the group. Where these were in conflict, our findings suggest that protesters accepted the security approaches collectively decided for the group. Group membership was conditioned on realising specific security goals related to the Anti-ELAB context -- anonymity in large public groups and confidentiality and authentication in small close-knit groups. Practices such as collective decision making to provide `security in numbers' and tactical `buy in' from group members substantiate the notion that, for the participants in our study, information security is a collective endeavour.

The idea of \emph{collectivity} in information security is not novel, yet, research on group-level information security is sparse -- and is largely limited to work on employee groups~\cite{AIS:JDHW19,CS:AlbHov09} and socialising contexts~\cite{CHI:WMKD20}. Moreover, usable security scholarship generally considers security at an individual level, as do user studies on messaging applications, see e.g.~\cite{AKPKKMTS17,SP:ASBDNS17,EUROSP:DNDS19,EWUS:SHWR16,SOUPS:VWOWHSZ17,SOUPS:VWOMCZ18}.
While, collectively, these studies highlight a series of usability shortcomings of messaging applications, they do not consider the social environment within which these are used, nor do they consider collective security practices which dominated our study. They generally treat such shortcomings as technological problems and/or incomplete mental models among individual users, rather than also considering how users' wider social context and collective, negotiated practices shape their use of these technologies and how (in)secure they feel in doing so.

Our findings demonstrate that the particularities of \emph{this} adversarial context, the Anti-ELAB protests, shaped participants' collective security needs and responses. Participants explained how social relations and trust were established at the protest sites rather than online and how this shaped their security practices, such as onboarding of new group members. In contrast to most usable security assumptions, our data shows that protesters go to great lengths to fulfil their security needs, conditioned on their adversarial setting and their group membership, but that such needs are not fulfilled by the technologies they rely on. 

As we show in Section~\ref{sec:high-risk-users}, other interview-based works on higher-risk users also emphasise the significance of the social context for the practice of information security. In bringing our findings into conversation with these studies, we note some high-level connections. For example, the participants in our study reported employing both technical and non-technical protection strategies, which has also been noted in recent studies on, e.g., journalists' use of security technology and related defensive practices~\cite{USENIX:MCHR15} and political activists' ``low tech'' protection mechanisms in the context of the Sudanese Revolution~\cite{SP:DSKB21}. Yet, while studies on other groups of higher-risk users, such as refugees and migrants, identify several cultural, social, economic and technological barriers that lead to unfulfilled security needs~\cite{CHI:GMSMTS18,SP:SLIRK18}, for the participants in our study, such barriers predominantly related to misconceptions about the security offered by the technology they relied on, the appropriation of insecure technology and their highly adversarial setting.

While it is possible to make some high-level connections between our findings and existing studies, the diversity of security concerns experienced by distinct groups and within specific contexts, requires grounded and situated research that is sensitive to this diversity. Moreover, our study, clearly illustrating how security is practised collectively among Anti-ELAB protesters, shows the critical need to situate technological security questions within the specific social contexts of groups, who share particular security goals. Thus, to understand collective security concerns and needs, future research should consider employing an ethnographic approach to ``unearth what the group (under study) takes for granted''~\cite[p.551]{PHG:Herbert00}.

\section{Conclusion}\label{sec:conclusion}

We conclude by summarising our key findings and by synthesising, with caution, requirements for (secure) messaging applications to serve the needs of protesters. Our interviews paint a diversified picture of group communication patterns, security needs and practices and they show how these are facilitated by a select few messaging applications and digital platforms.

Protesters rely heavily on Telegram and WhatsApp for their communication. Our findings illustrate how central these tools are for organising on the ground, by facilitating a collective approach to establish tactics, e.g.~through anonymous polls, which was seen to provide both `security in numbers' and `buy in' for the chosen tactic. These decisions were made in groups of varying size and the administrators of these groups adopted the roles of leaders in these `leaderless' and `decentralised' protests. Overall, we found that these protests were organised in a mix of large public and small close-knit groups, with differing security requirements: anonymity within the group, on the one hand, and confidentiality and authentication, on the other. To bridge the conflicting requirements of anonymity and trust, participants reported a long, offline onboarding process before adding new members to a group.

The participants in our study developed tactics to detect compromise and to achieve some form of forward secrecy, i.e.~protection of secrets against a later compromise. Group members monitored the movements of fellow group members to eliminate traces of the group chat from their phone in case of an arrest and to render legal aid. This explains the importance attributed to the ability to remotely delete messages on other people's devices. Participants adopted a variety of practices to address (perceived) shortcomings of digital communications and conflicting security needs. For example, to facilitate pseudonymity, compartmentalisation through the use of multiple devices and burner phones was widespread.
Participants also reported how security decisions were collective, requiring group members to buy into the security practices of their group. This was a process fraught with conflict as differing security needs confronted each other.

For designers, several requirements on (secure) messaging applications emerge from our data: support for both (small) private and (large) public groups, the avoidance of phone numbers or other personally identifiable information and the ability of administrators to control messages and participation in groups. In particular, there is a clear distinction in security requirements for different types of groups: anonymity in large groups, confidentiality up to forward secrecy in small groups. In addition, going beyond strictly messaging, several features such as polls and live location sharing emerged as key enablers for participants. Participants also expressed a strong desire to be able to have control over their messages after sending them, such as on-demand remote message deletion.

However, we caution against taking this list of requirements as a blueprint. First, our data only covers interviews with 11 participants. Second, these feature requests are informed by what existing technologies provide and thus do not necessarily represent the horizon of what is possible or desirable. Third, as we discuss above, the security guarantees provided by some of the employed tactics, particularly remote message deletion, are limited. Fourth, our data presents information security as a negotiated, conflict-laden and changing practice, suggesting that a universal solution may not exist.

\section*{Acknowledgements}

We thank the participants for speaking to us and the gatekeepers for their assistance in establishing contact with participants. The research of Marekov\'{a} was supported by the EPSRC and the UK Government as part of the Centre for Doctoral Training in Cyber Security at Royal Holloway, University of London (EP/P009301/1).

{
\fontsize{9.3}{11}\selectfont
\bibliographystyle{splncs03}
\bibliography{bibexport.bib}
}

\fullversion{
\newpage
\appendix

\section{Interview topic guide}\label{sec:interview-topic-guide}

Topic guide used for semi-structured interviews with individuals who have been involved in the Anti-Extradition Law protests in Hong Kong (HK). While structured around five key topics, it includes prompts, examples and follow-on questions to guide the interview.

\subsection*{Topic 1: The use of communication technology in HK}

The aim of this topic is to establish existing communication patterns in HK, beyond the protests, before focusing on the protest context in subsequent topics. 

\begin{itemize}
\item Preferred mode of communication in HK?
\item Popular online platforms in HK?
\item Why do you think they are popular in HK? 
\item Why use these online platforms?
\item Benefits/disadvantages?
\item Use of large group chats/forums in HK? 
\item Why? Why not?
\item The use of online platforms by HK authorities in everyday communications?
\end{itemize}

\subsection*{Topic 2: Platforms and group chats/forums in the Anti-ELAB protests}

This topic focuses on how the protest context changes communication patterns, if at all. 

\begin{itemize}
\item How does the use of online platforms change during protests?
\item To what extent are some platforms used more than others? Which? 
\item Why do you think that is? Examples?
\item How do communication patterns change during protest? In terms of: networks, group chat/forum size, frequency? 
\item Where are online platforms used? In the street?
\item At what points in the protests are online platforms relied upon? By whom? Why?
\item For what purpose are platforms/group chats/forums used? For planning and organisation? Data gathering? Verification of rumours?
\item Temporal aspects: to what extent do the dynamics of the protests reflect app use/avoidance?
\item Are you aware of Bluetooth enabled applications being used? Bridgefy? Concerns about security shape communication technology use during the Hong Kong protests
\end{itemize}

\subsection*{Topic 3: How concerns about security shape communication technology use during the Anti-ELAB protests}

This topic focuses specifically on concerns related to the use of communication technology during the Anti-ELAB protests. 

\begin{itemize}
\item Concerns about online communication switch-off? Likelihood?
\item What would be the concern?  
\item Who would be concerned? Protesters? Why?
\item How would a switch off affect the protests?
\item Concerns about infiltration of specific applications?
\item To what extent do people speak more openly on one app over the other? Specific applications? Why?
\item Concerns about information shared? Why? Examples?
\item Concerns about information received? Why? Examples?	
\end{itemize}

\subsection*{Topic 4: Notions of security within online/offline networks during the Anti-ELAB protests}

This topic focuses on how networks -- online and offline -- are shaped by different notions of security.

\begin{itemize}
\item To what extent do people know participants in their group chats/forums? 
\item How do these groups map onto offline groupings?
\item How are people added and removed from networks? Platform specific? Group chat/forum specific? Specific processes of authentication? 
\item To what extent do online and offline onboarding map onto each other?
\item What are the main disruptive factors within online networks?
\item Concerns about being seen to be present in protest related chat groups? Why? Examples?
\item Wider networks: what repercussions might protesters fear? Affecting themselves, their family, their friends etc.? 
\item Who might they fear repercussions from?
\end{itemize}

\subsection*{Topic 5: Designing secure communication platforms for high-risk environments}

As a ``wrap-up'', this topic explores future directions in the design of secure communication technology for high-risk contexts.

\begin{itemize}
\item What should designers of secure communication platforms design for based on your experience with these protests? Why?
\end{itemize}
}{}

\end{document}